УДК 004.9 + 520.84

# ЗВЕЗДЫ КАТАЛОГА GTSh10 В «ПЛАНЕ ШАЙНА»

*М.А. Горбунов[1], А.А. Шляпников[2]*

[1] ФГБУН «Крымская астрофизическая обсерватория РАН», Научный, Крым, Россия, 98409
*mag@craocrimea.ru*
[2] ФГБУН «Крымская астрофизическая обсерватория РАН», Научный, Крым, Россия, 98409
*aas@craocrimea.ru*

**Аннотация.** С целью пополнения базы данных фотометрических и спектральных наблюдений красных карликовых звёзд каталога GTSh10 проверено наличие этих объектов среди исследованных в рамках выполнения «Плана академика Г.А. Шайна» (далее – «План Шайна»). Рассмотрен вопрос перекрестной идентификации данных интерактивной версии GTSh10 и каталогов «Плана Шайна», а также доступа к информации архива фотографических наблюдений, на основе которых они были составлены. Статья проиллюстрирована примерами работы с полученным списком средствами виртуальной обсерватории с целью анализа состояния объектов и/или уточнения их фотометрических и спектральных характеристик в середине прошлого века.

STARS FROM THE GTSh10 CATALOGUE IN "SHAJN'S PLAN", by M.A. Gorbunov, A.A. Shlyapnikov. We examined objects from the GTSh10 catalogue among the studied stars as a part of implementing the 'Plan of Academician G.A. Shajn" (hereinafter - "Shajn's Plan") in order to replenish the database of photometric and spectral observations of red dwarfs. We consider the question concerning the cross-identification of data from the GTSh10 interactive version and the "Shajn's Plan "catalogs, as well as the access to the archive of photographic observations. The article is illustrated with examples of working with the obtained list by means of the Virtual Observatory in order to analyze the state of objects and/or to refine their photometric and spectral characteristics in the middle of the past century.

**Ключевые слова:** каталоги

## 1 Введение

Создание интегрированной структуры базы данных, составляющей основу проекта «Крымская астрономическая виртуальная обсерватория» (Шляпников, 2007, 2013), предполагает определение связей накопленной в обсерватории информации. Это и оригинальные наблюдения, сохраняемые в стеклянной библиотеке и полученные уже в цифровом формате, и опубликованные результаты исследований, в том числе представленные в виде каталогов.

В 2010 году был подготовлен каталог GTSh10, содержащий 5535 объектов, в основном звёзд карликов нижней части Главной последовательности. Подробное описание этого Каталога дано в первом выпуске 107 тома «Известий Крымской астрофизической обсерватории» (Гершберг и др., 2011).

14 каталогов, содержащих информацию о звёздных величинах, показателях цвета и спектральных классах ~ 35000 звёзд, были получены реализации «Плана академика Г.А. Шайна» по изучению структуры Галактики. Они опубликованы в 9-ти томах «Известий Крымской астрофизической обсерватории» с 1953 по 1963 годы, в 7-м томе Трудов Рижской астрофизической лаборатории 1958 года, в 7-м томе Бюллетеня Вильнюсской астрономической обсерватории 1963 года и в 136-м томе Сообщений ГАИШ в 1964 году (Pronik, 2005). В 2007 году начат перевод каталогов в машиночитаемый формат (Горбунов, Шляпников, 2017а, 2017б). Рисунок 1 иллюстрирует распределение на небесной сфере в галактической системе координат объектов из каталога GTSh10 и области покрытия закаталогизированными объектами из «Плана Шайна». Отметим, что в некоторых случаях, создание каталогов было ориентировано на выявление звёзд ранних спектральных классов, возбуждающих свечение туманностей. Однако большая часть каталогов содержит практически все объекты доступные для исследований в выбранных областях.

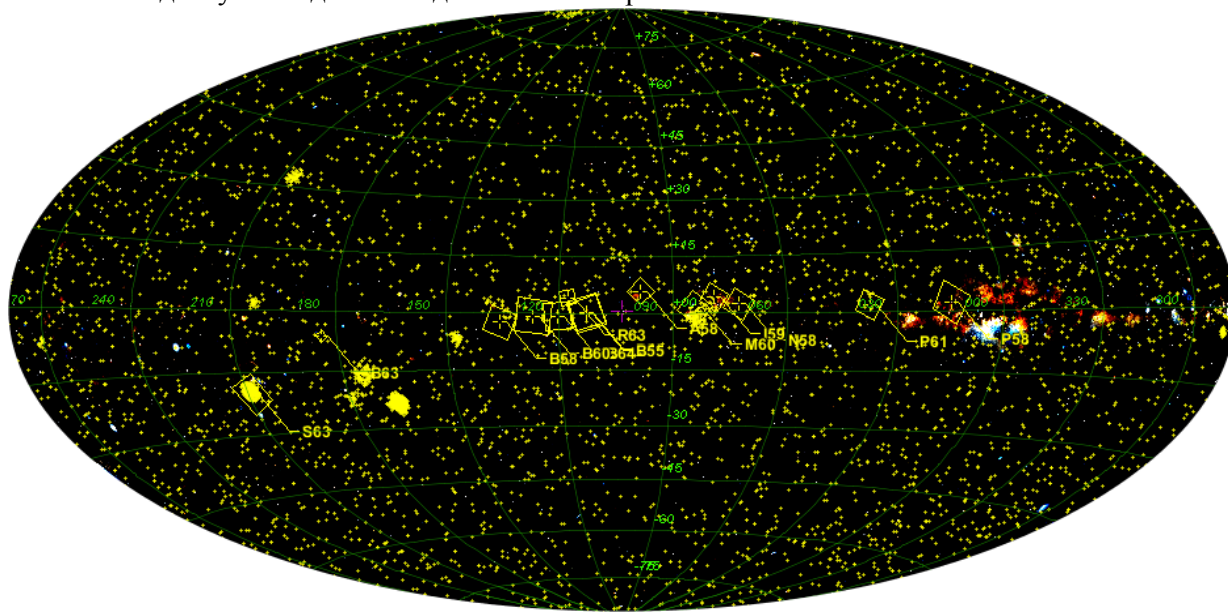

Рис. 1. Распределение на небесной сфере объектов из каталога GTSh10 и области покрытия закаталогизированными объектами из «Плана Шайна» (в галактической системе координат).

Основой для создания каталогов «Плана Шайна» стали наблюдения, выполненные в Крымской астрофизической обсерватории. Это 800 прямых и 500 снимков, полученных с объективной призмой. Учитывая значительную площадь покрытия небесной сферы вдоль Млечного Пути (более 1300 квадратных градусов), особый интерес представляют переменные звёзды, в частности с иррегулярными изменениями блеска, проявлениями вспышечной активности и другие пекулярные объекты. Подробно об архиве наблюдений, перспективах его оцифровывания и использования для решения астрофизических задач можно ознакомиться в ряде публикаций (Bondar', 2002, Bondar', Rumyantsev, Shlyapnikov, 2005, Bondar', Shlyapnikov, 2006, Gorbunov, Shlyapnikov, 2013, Pakuliak et al., 2014).

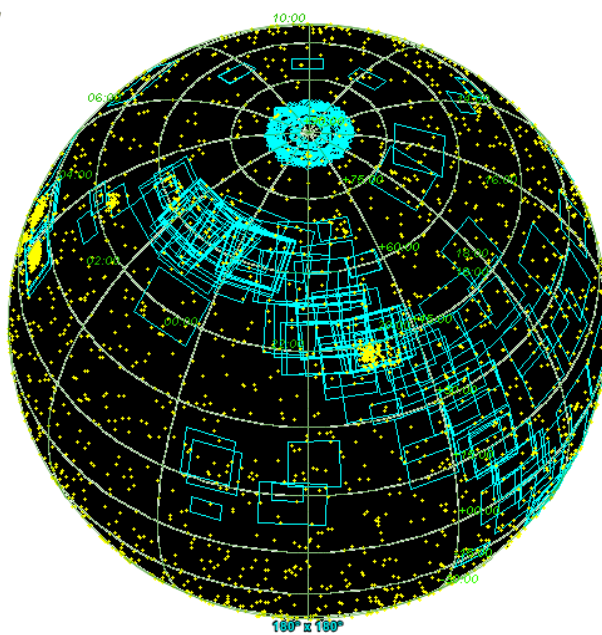

Рис. 2. Распределение на небесной сфере объектов из каталога GTSh10 и области покрытия негативами, полученными по «Плану Шайна» (в экваториальной системе координат).

По первому и второму рисунку видно, что значительное число звёзд из GTSh10 должно присутствовать, как в каталогах, так и на негативах «Плана Шайна». Отметим, что эти каталоги содержат, как правило, фотометрию, выполненную в двух полосах, и спектральную классификацию звёзд. Информация каталогов и возможность независимого определения блеска звёзд из GTSh10 по негативам, позволят оценить состояние объектов на момент получения данных.

### 2 Идентификация звёзд GTSh10 в каталогах «Плана Шайна»

Проблемы интерактивного использования каталогов, созданных по «Плану Шайна», подробно рассмотрены в одной из статей, посвящённой их преобразованию в цифровой формат (Горбунов, Шляпников, 2017б).

Для перекрёстной идентификации объектов GTSh10 и звёзд из каталогов «Плана Шайна» применялся интерактивный атлас неба Aladin (Бонреаль и др., 2000). Таблица 1 иллюстрирует звезды из каталога GTSh10, найденные среди объектов из каталога B58 (Бродская, Шайн, 1958). В колонках указаны следующие данные. Координаты по обоим каталогам R.A.$_{GTSh10}$, Decl.$_{GTSh10}$ и R.A.$_{B58}$, Decl.$_{B58}$ Номер по каталогу GTSh10 и обозначение в базе данных SIMBAD (№ GTSh10 и Name SINBAD). Звездная величина в соответствующей полосе из GTSh10 и B58 (mag и B$_{B58}$). Спектральные классы по GTSh10 и B58 (Sp и Sp$_{B58}$). Наличие оптических вспышек OF. Номер по каталогу BD и показатель цвета B-V по B58.

Таблица 1

| R.A. GTSh10 | Decl. GTSh10 | № GTSh10 | Name SIBAD | mag | Sp | OF | R.A. B58 | Decl. B58 | BD | Sp B58 | B B58 | B-V B58 |
|---|---|---|---|---|---|---|---|---|---|---|---|---|
| 02 14 44.40 | +59 47 57.0 | 420 | V* V603 Cas | 11.18 B | M 0.5 | F | 02 14 44.37 | +59 47 56.6 | +59 151 | F0 | 11.29 | - |
| 02 13 31.60 | +60 26 03.0 | 415 | V* V601 Cas | 11.71 B | - | F | 02 13 31.62 | +60 26 03.6 | +60 77 | G0 | 11.66 | - |
| 02 30 39.60 | +61 00 25.0 | 467 | V* V612 Cas | 12.50 B | M 2 | F | 02 30 39.63 | +61 00 25.2 | +60 192 | K0 | 12.03 | - |
| 02 55 56.90 | +61 31 16.0 | 552 | HR 860 | 5.60 | F4/8 | - | 02 55 56.75 | +61 31 15.8 | +61 430 | F5 | (6.0) | - |
| 02 22 26.30 | +61 35 35.0 | 457 | V* V607 Cas | 12.10 B | M | F | 02 22 26.27 | +61 35 34.9 | +61 126 | B5 | 12.18 | 0.52 |
| 02 24 52.90 | +61 53 47.0 | 459 | V* V609 Cas | 12.30 B | M 3 | F | 02 24 52.60 | +61 53 46.0 | +61 134 | K5: | 12.05 | - |

### 3 Звёзды GTSh10 в архиве негативов «Плана Шайна»

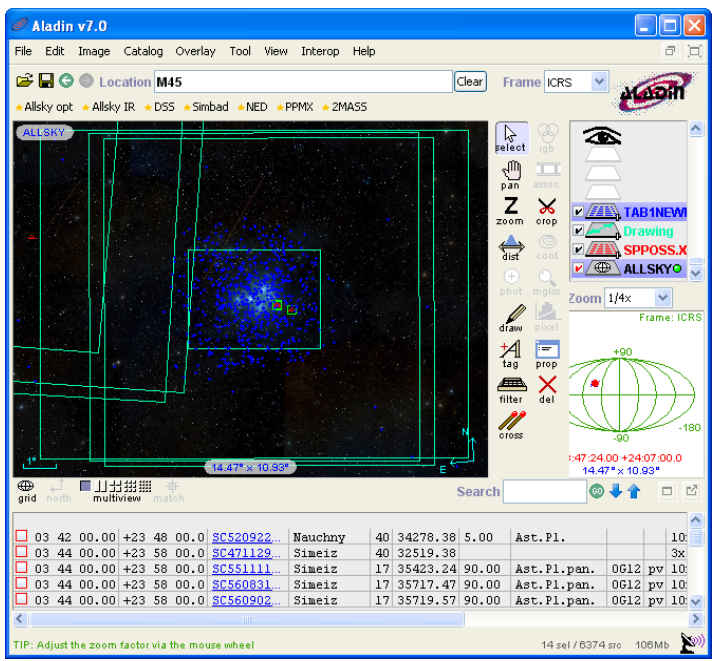

Как было отмечено во введении, при выполнении «Плана Шайна» был получен наблюдательный материал, представляющий интерес для анализа состояния объектов на значительных интервалах времени. Так как данная работа проводилась с целью отработки методики фотометрии звёзд из GTSh10 по архивным наблюдениям, для большей наглядности была выбрана область, содержащая значительное число объектов на единицу площади отснятого неба. В данном случае – это область рассеянного скопления M45 (Плеяды). На рис. 3 показан интерфейс Aladin с 472 звёздами из GTSh10 и негативами из стеклянной библиотеки, на которых зарегистрирована данная область (их оказалось 14 в коллекции «План Шайна»).

Рис. 3. Область Плеяд со звёздами из GTSh10 и контурами негативов из архива.

Далее негативы были просмотрены на предмет их качества. Для этого последовательно каждый из 14 негативов автоматически пересылался в новое окно Aladin по гиперссылке, указанной в нижней части интерфейса и обозначающей номер изображения в архиве. В результате были отобраны негативы SC570202_42_2 и SC560902_42_4 (рис. 4). Следующим шагом стала обработка негативов с помощью программы Sextractor (Бертин, 1996). Обнаруженные объекты были откалиброваны в системе каталога Tycho-2 (Хог, 2000) (рис. 5). Линеаризация характеристических кривых на данном этапе не производилась.

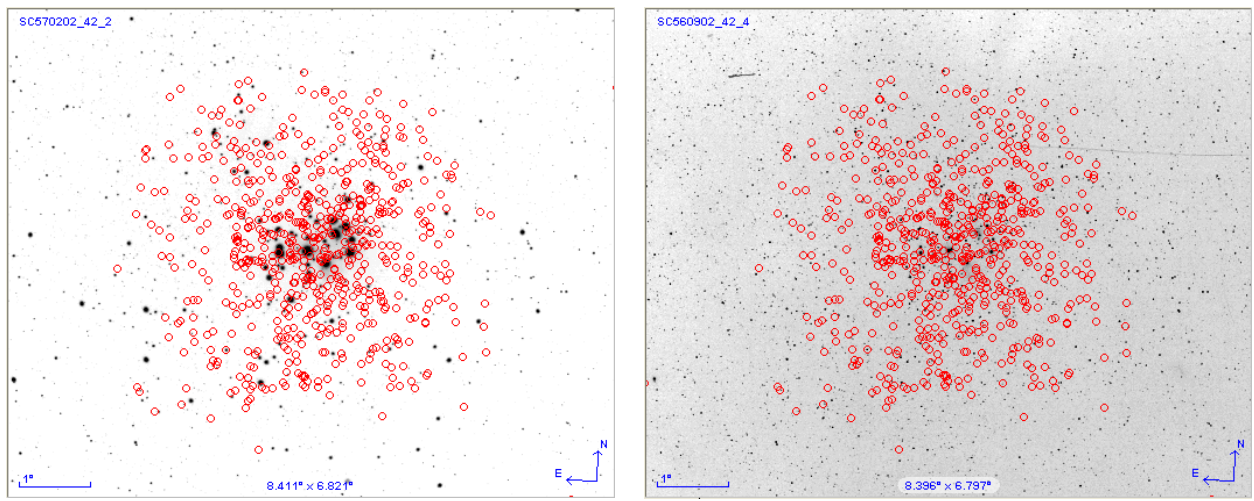

Рис. 4. Отобранные негативы, с указанием положений звёзд из GTSh10.

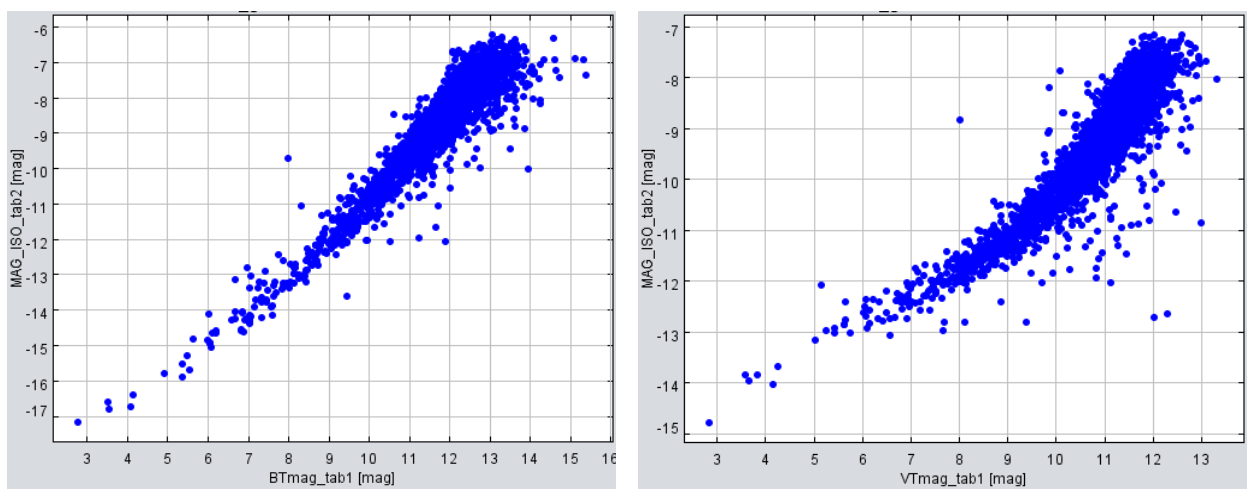

Рис. 5. Калибровочные кривые негативов.

Отметим, что при обработке использовались уменьшенные (превью) изображения, доступные on-line. Поэтому предъявлять высокие требования к фотометрической точности при оценках блеска объектов по этим негативам не логично, прежде всего, т.к. размер звёзд на нём в угловой мере составляет около 1ʺ и для корректной калибровки изображений необходимо производить суммирование блеска объекта вблизи калибруемого с такой диафрагмой. Тем не менее, рисунок 6 показывает хорошее согласие между звёздными величинами V по GTSh10 и определенным по негативу SC560902_42_4.

## 4 Заключение

Рассмотренная в постере процедура поиска объектов каталога GTSh10 по данным «Плана Шайна» позволяет пополнить базу данных фотометрических и спектральных наблюдений красных карликовых КрАО. Возможность работы с интерактивными приложениями Международной виртуальной обсерватории обеспечивает доступ к мировым астрономическим базам данных.

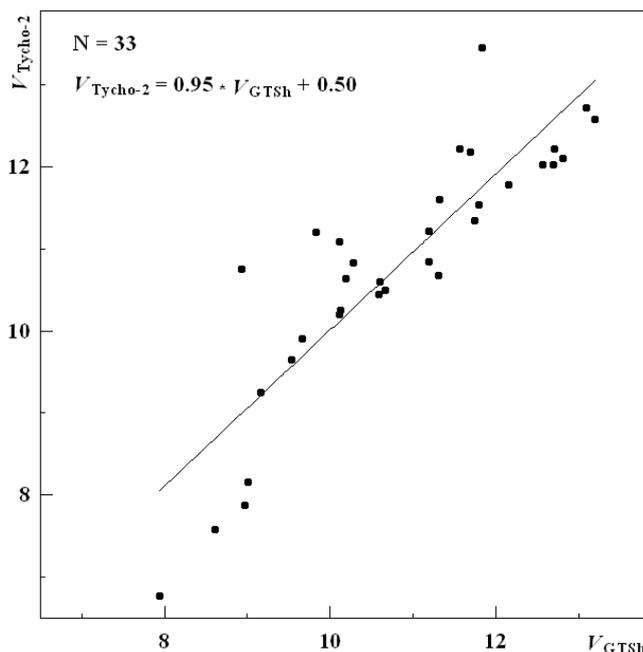

Рис. 6. Сравнение звёздных величин V по GTSh10 и определенных по негативу SC560902_42_4.